\documentstyle[aps]{revtex}

\def\be{\begin{equation}}
\def\ee{\end{equation}}
\def\ba{\begin{eqnarray}}
\def\ea{\end{eqnarray}}
\def\l{\label}
\def\n{\nonumber\\}
\def\g{\gamma}
\def\D{\Delta}
\def\a{\alpha_s}

\begin{document}

\title{Unifying approach to hard diffraction
}

\author{H. Navelet and R. Peschanski\thanks{%
CEA, Service de Physique Theorique, CE-Saclay, F-91191 Gif-sur-Yvette Cedex,
France}}
\maketitle

\begin{abstract}
We find a consistency between two different approaches of hard diffraction, 
namely the QCD dipole model and  
the Soft Colour Interaction approach.   A theoretical 
interpretation in terms  of S-Matrix and 
perturbative QCD  properties in the small $x_{Bj}$ regime is proposed.
\end{abstract}

\bigskip

{\bf 1.} In the present paper, we  focus on two different 
theoretical 
approaches 
to hard diffraction,  which are shown to be compatible and complementary.  The 
first  one 
 is based on an extension \cite {bi96} to hard diffractive 
processes of the ``QCD dipole'' 
approach in the small $x_{Bj}$ regime of perturbative QCD. In this picture,  the 
hard photon  is supposed to probe the parton structure of the Pomeron 
considered as an hadronic particle. In the original 
Ingelman-Schlein formulation \cite 
{in85}, this hard probe was formulated in the parton model, later complemented 
by QCD evolution equations \cite{blu}. In the dipole model, 
this hard interaction is  described via  the Balitsky Fadin Kuraev Lipatov 
(BFKL) resummation \cite{bfkl}. A major uncertainty for this model is the 
relative 
normalization of diffractive over non-diffractive cross-sections which stays 
beyond the perturbative framework. 

 A second approach \cite{in96} is 
the Soft Colour Interaction  (SCI) model, 
where hard diffraction is described as the superposition of two processes. At 
 short time,  the hard probe initiates a typical 
deep-inelastic interaction with  colour quantum numbers exchange. Then,   at 
large 
times/distances, a ``soft'' colour 
interaction 
is assumed to rearrange the colour quantum numbers and gives rise to singlet 
exhanges -and thus diffraction- with a probability of order ${\bf \frac 
1{N_c^2}},$ where $N_c$ is the number of colours\footnote{More recent versions 
of the model consider a probability factor of the same order but not necessarily 
connected with colour.}. While in this approach, the relative normalization is 
fixed, the exact nature of the interplay between soft and hard components is not 
known.

In the present paper, we propose a way to relate  these two approaches which 
allows one  to consider them as compatible and 
complementary. This   provides a definite prediction for both the normalization 
and the analytic form of the amplitude. Our main results are the following:

 {\bf i)} The interplay between hard and soft components of hard diffraction is 
expressed via ``effective'' parameters of 
a Pomeron interaction determined from  
leading 
log perturbative  QCD 
resummation. It is found 
to depend not only on  $Q^2$ but also on the ratio
$\eta = 
 \left(Y-y\right) /y,$ where Y (resp. y) are the total (resp. gap) rapidity 
interval.

We obtain
\begin{equation}
F_{T,L}^{Diff}(Q^2,Y,y) = {\bf \frac 1{N_c^2}}\ 
\frac {{\cal N}^{tot}}{x_P}  
\ \frac {e^{2y\D}}{4\pi \D ^{''} y}
\ \sqrt {\frac 2
{1\!+\!2\eta}}
\exp \left\{ (Y\!-\!y)\ {\bf \epsilon}   _s \right\}\ \left(\frac 
Q{Q_0}\right)^{2{\bf \g_s}}
\exp{\left( - \frac {2\log^2\!\left(\frac 
Q{Q_0}\right)}{D_s 
(Y\!-\!y)}\right)} \ ,
\l{biczy10}
\end{equation}
with
\begin{equation}
\D(x) \equiv \frac {\a 
N_c}{\pi}\left\{2\psi(1)-\psi(x)-\psi(1-x)\right\}
\  \sim \D + \frac 
{\ \D^{''}}2 \left(1/2\!-\!x \right)^2
\l{chi}
\end{equation}
 is the BFKL  evolution kernel \cite{bfkl} 
(together with its gaussian approximation near 
the minimum at $x=1/2$) and $Q_0$ a 
non-perturbative scale associated with the proton. We have
\begin{equation}
{\bf \g_s}= \frac {\eta}{1+2\eta}\ ;\ 
{\bf \epsilon}_s  =\D + \frac {\D^{''}}{8 (1+2\eta)}\ ;\ D_s = \frac 
{1+2\eta}{\eta}\ \D^{''}\ .
\l{epsilon} 
\end{equation} 

Note that one may write $
F_{T,L}^{Diff} \ \sim
\ \sigma^{tot}_{\gamma^*-P}\times e^{2y\Delta }/(4\pi \D ^{''} y) ,$ which is 
the 
known ``triple Pomeron'' formula \cite{ka79}
where 
$\sigma^{tot}_{\gamma^*\!-\!P}$  defines the effective interaction cross-section 
of a virtual photon with a  BFKL Pomeron $e^{y\Delta }/\sqrt {4\pi \D ^{''} y}$ 
derived from the QCD dipole formalism. By analogy  with BFKL \cite {bfkl}, 
$\g_s, \epsilon_s$ and $D_s$ 
can be defined, respectively, as  the  anomalous
dimension,
intercept and diffusion parameter  of an ``effective'' BFKL $\gamma^*\!-\!P$ 
cross-section.

The normalization, which remains unknown
in the QCD dipole model description, is  determined as
  the product of the factor ${\bf \frac 1{N_c^2}}$ and the  
normalization  factor ${\cal N}^{tot}$ of the non-diffractive structure 
function, 
according 
to the SCI prescription, see below.

{\bf ii)}
In SCI models,  the following relation  between 
the total structure 
function and the overall contribution of hard diffraction  at 
fixed value reads
of $x_{Bj}:$

\begin{equation}
F_{T,L}^{Diff/tot}\!\! \equiv \!\int_{x_{Bj}}^{x_{gap}} dx_P \ F_{T,L}^{Diff} = 
{\bf \frac 1{N_c^2}}\ 
F_{T,L}^{tot}
\l{biczy4}
\end{equation}
where $\log 1/{x_{gap}}$ is the 
minimal  rapidity 
gap. If we insert the QCD dipole prediction for $F_{T,L}^{Diff}$ in formula 
(\ref{biczy4}), we find
\begin{equation}
F_{L,T}^{tot} = \frac {{\cal N}}{\bf {N_c^2}}\left(\frac 
Q{Q_0}\right)^{2\g^{*}}
\frac {\exp \left(Y\D (\g^{*})\right)}{\sqrt {2\pi\D ^{''} \ Y}}  \ ,
\l{tot1}
\end{equation}
\noindent
which is equivalent to a canonical  BFKL expression for 
non-diffractive structure functions, apart the substitution of the BFKL 
effective anomalous dimension $\g_{BFKL}$ by $\g^{*},$ namely
\begin{equation}
\g_{BFKL} = \  1/2-  2\frac 
{\log\left(\frac Q{Q_0}\right)}{\D ^{''} \ Y}
\ \rightarrow \ \g^{*} = cst. \sim 0.175\ ,
\l{value}
\end{equation}
where the ``universal'' value $\g^{*}$ is  solution (for $0<\g<1/2$) of the 
implicit equation $
2\D\left(\frac{1\!-\!\g}{2}\right)-\D(\g)=0\ $.
Here, it is interesting to note that the shift (\ref{value}) may be useful to 
avoid the 
objection to SCI models \cite {dok} based on ``Low's theorem''  following which 
soft 
colour radiation cannot be emmitted from inside a partonic  process. The 
differences we find with the original model means that, in our dipole 
formulation, the soft colour interaction indeed seems to modify  the initial 
parton kinematics.  
 
 {\bf iii)} Using S-Matrix properties of triple-Regge contributions, a relation 
is 
found 
between  discontinuities of a $3 \to 3$ amplitude and the  two approaches to 
hard 
diffraction we consider.
Following old results of S-Matrix theory in the Regge domain \cite {mu}, and as 
sketched in 
Fig.1, one may 
consider three types of discontinuities of a $3 \to 3$ amplitude representing 
hard diffraction. A single discontinuity over the diffractive $mass^2$ 
describing the hard  Pomeron interaction, 
a double discontinuity  taking into account the 
analytic discontinuity  in the subenergy variable of one of the incident
 Pomeron exchanges (and its complex conjugate) for the SCI 
model
and the full  triple discontinuity including those of
the two 
Pomeron 
exchanges, which is characteristic of the QCD dipole model description 
\cite{bi96}. An interesting new feature is thus the S-Matrix interpretation of 
the 
SCI approach  as a 
specific 
double discontinuity of the $3 \to 3$ forward amplitude, which formulates the 
model in terms of simultaneous exchanges of a soft and a hard Pomeron.

\bigskip

{\bf 2.} Let us sketch the derivation of our results.

Our  starting point   is a triple-Regge  formula for the diffractive
 structure function 
 for longitudinal and 
transverse photon in the QCD dipole formalism:
\be
F_{T,L}^{Diff}(Q^2,Y,y)\sim \frac {{\cal N}^{Diff}}{x_P}
\int_{c-i\infty}^{c+i\infty}\frac 
{d\g_1}{2i\pi}\frac {d\g_2}{2i\pi}\frac {d\g}{2i\pi}\ \delta 
(1\!-\!\g_1\!-\!\g_2\!-\!\g)\ 
  \left(\frac Q{Q_0}\right)^{2\g}\ \exp \left\{y (\D(\g_1)+\D(\g_2)) + 
(Y\!-\!y)\ 
\D(\g)\right\}\ ,
\l{biczy}
\ee
where
$\D(\g)$ 
 is the BFKL evolution kernel (\ref{chi}) and ${\cal N}^{Diff}$ is a 
normalization containing both QCD perturbative and non-perturbative factors 
\cite {ba96}. Strictly speaking \cite {ba96} the $\delta$-function is the 
unique contribution  in the differential diffractive 
structure function at momentum transfer $t=0.$ However, it can be shown  
that this is the dominant perturbative contribution even at non zero transfer 
due to specific properties \cite{na02} of  the analytic QCD triple-Pomeron 
couplings \cite{na01} .

The first step of the computation of formula (\ref{biczy}) is 
to use the 
saddle-point 
approximation \cite{ba96}  
at large $y$ to integrate over the difference $\g_1\!-\!\g_2.$ One easily gets
\be
F_{T,L}^{Diff}= 
{\cal N}^{Diff}\frac 1{x_P }\ \int_{c-i\infty}^{c+i\infty}\frac 
{d\g}{2i\pi}\sqrt {\frac 
1 {4\pi \D ^{''}\! (\frac {1-\g} 2)\ 
y}} 
\exp \left\{2y\ \D\left(\frac {1\!-\!\g} 2\right) + (Y\!-\!y)\ \D(\g) + 2\g 
\log\frac 
Q{Q_0}\right\} \ .
\l{biczy1}
\ee

Using the  gaussian approximation (\ref{chi}) for the BFKL kernels $\D(\g)$ and 
$\D\left(\frac {1\!-\!\g} 2\right)$ in the relevant interval $0 <\g<1/2,$  and 
again a saddle-point 
approximation at 
 large rapidity gap $y$, one obtains, up to a 
normalization factor,  formula (\ref{biczy10}), with  $\g_s, \epsilon_s$ and 
$D_s$  defined as in (\ref{epsilon}). The normalization ${\cal N}^{Diff}$ is 
not yet specified at this stage.
 
 The derivation of the normalization is coming from the comparison with the 
SCI approach. Inserting  (\ref{biczy}) in the integral  
(\ref{biczy4}), one is led to perform a two-dimensional  saddle-point 
approximation in the $y,\g$ complex plane. The saddle-point equations read:
\ba
-y\ \D^{'}\left(\frac 
{1\!-\!\g} 2\right) + (Y\!-\!y)\ \D^{'}(\g) + 2
\log\frac 
Q{Q_0}=0
\n
2\D\left(\frac{1-\g}{2}\right)-\D(\g)=0\ ,
\l{cols}
\ea 
whose solution $(y^*,\g^*)$ is
\ba
y^* &=& \left(Y+\frac {2\log\frac Q{Q_0}}{\Delta^{'}(\g^*)}\right)
\ \left(1+\frac {\Delta^{'}\left(\frac{1-\g^*}{2}\right)}
{\Delta^{'}(\g^*)}\right)^{-1} 
\n
\D(\g^*)&=&2\D\left(\frac{1-\g^*}{2}\right)\ \,
\l{solution}
\ea
resulting in a value of $\g^*\simeq 0.175 ,$   which is ``universal'', i.e. 
independent of the 
kinematics of the reaction.

After  computation of the prefactors to the 
saddle-point approximation, one finds:
 \be
F_{T,L}^{Diff/tot}\! =\!   
{\cal N}^{Diff}
\frac 1
{\vert\D^{'}(\frac {1\!-\!\g^*}2)\!+\!\D^{'}(\g^*)\vert}
\left(\frac Q{Q_0}\right)^{2\g^*}  \frac{\exp \left(Y 
\D(\g^*)\right)}{\sqrt{4\pi\D^{''}\left(\frac{1-\g^*}{2}\right)y^*}}
\ .
\l{biczy7}
\ee
Note that it is the linearity in $y$ of the saddle-point equation (\ref{cols}) 
which allows one
to get such an elegant form for the integrals.
Using (\ref{biczy7}) and by comparison with the canonical BFKL formula we 
identify the ``hard'' component of the SCI model by  the substitution $
\g_{BFKL} \rightarrow  \g^{*} ,$ see  
(\ref{value}). Then using the SCI ansatz (\ref{biczy4}), we 
obtain the relation
\be
{\cal N}^{Diff} \approx \frac {{\cal N}^{tot}}{\bf {N_c^2}}\times
\ {\vert\D^{'}(\frac {1\!-\!\g^*}2)\!+\!\D^{'}(\g^*)\vert}
\l{norma}
\ee
which fixes the relative  normalization of the diffractive vs. non diffractive 
structure functions. This ends the derivation of formulae 
(\ref{biczy10}-\ref{value}).

\bigskip

{\bf 3.} Let us finally come to the S-Matrix interpretation of our approach.  
For every fixed but arbitrary value of the parameters $\g,\g_1,\g_2,$ the 
triple-Regge formula (\ref{biczy}) can be obtained from the canonical 
formulalism\footnote{These triple Regge formulae, relevant for {\it inclusive 
cross-sections}, have to be distinguished from the AGK cutting rules which are 
relevant for the two and multi Pomeron contributions to the $2 \to 2$ {\it 
elastic} amplitudes in both the S-Matrix framework  \cite{AGK} and perturbative 
QCD at leading log level \cite{AGK1}.} 
corresponding to the vertex of three Regge pole singularities\footnote{We 
neglect at this stage the complications due to Regge cuts or other more 
sophisticated singularities and deal with simple effective Regge poles.} in the 
complex plane of angular momentum \cite{ka79}. As such, one can  make use of the 
  important S-Matrix  
Mueller-Regge relation \cite {mu}, valid in kinematical regions including  the 
triple-Regge limit,
 between 
  semi-inclusive  amplitudes and  specific 
discontinuity contributions of forward elastic $3 \to 3$ amplitudes. It 
naturally applies to hard 
diffraction 
initiated by a virtual photon, as sketched in Fig.1,  namely 
\be
\gamma^* + p\to p + X \ \iff Disc_1\left\{\gamma^* \bar p \ p \ \to 
\gamma^*  \bar p\ p  
\right\} \ 
.
\label{3to3}
\ee
Quite interestingly, the existence of Regge phase factors
allows one  to  relate 
other discontinuities of $A(3 \to 3)$ to $Disc_1 A.$   As sketched in 
Fig.1, one may also
consider a double discontinuity $Disc_2 A(3 \to 3)$ taking into account also the 
analytic discontinuity  in the subenergy of one of the incident
 Pomeron exchanges 
and a triple discontinuity $Disc_3 A(3 \to 3)$ including the discontinuity 
over the two 
Pomeron 
exchanges. The 
expression of the 
discontinuities, through generalized unitarity relations, is obtained through 
the 
imaginary part of the 
relevant Regge phase 
factors \cite{mu}. Moreover, one finds an equality relation
$Disc_1 A \! = \!Disc_2 A\! = \!Disc_3 A $ which is due to the fact that the 
discontinuity  taken over the mass variable (corresponding to diffractively 
produced states) is common to all three cases in Fig.1 and factorizes 
the same
 $p\bar p$ vertex in 
$A(3\! \to \!3)$ (cf. the classical derivation in the last paper of 
Ref.\cite{mu}). 

Let us now take 
advantage of the hard probe in the process, allowing one to introduce in the 
game the 
(resummed) 
perturbative QCD expansion at 
high energy (small $x_{Bj}$).   In a generic S-Matrix approach, the 
analytic discontinuities of scattering amplitudes are related to a 
summation over a complete set of asymptotic {\it hadronic} 
final states. If however, 
the underlying microscopic field theory is at work with small renormalized 
coupling constant due to 
the hard probe, it is possible in some cases to approximate the same 
discontinuity using a 
complete set of {\it 
partonic} states. In particular, at high energy and within the approximation of 
leading logs (and also 
large $N_c$), QCD dipoles can be identified  as providing such a basis \cite 
{bi96}. 

The discontinuity $Disc_3$ appears naturally in the dipole formulation of hard 
diffraction \cite{bi96}. Indeed, hard diffraction calculations make use of the  
probability distribution for finding two dipoles in the wave-function of one 
initial dipole in the  virtual photon through the exchange over a
 BFKL Pomeron exchange described by its discontinuity in the $Y-y$ 
rapidity range. Then,  each of these dipoles interact with the target  through  
two perturbative (BFKL) Pomeron exchanges described by their own discontinuities 
over the $y$ range. Thus the calculation of scattering amplitudes implies the 
full 
discontinuity over three Pomerons described by  intermediate dipole 
interactions, which is nothing but 
$Disc_3.$

The appearance of  $Disc_2$ is natural in the Regge formalism \cite{mu}. However 
a physical interpretation involving perturbative QCD contributions has not been 
previously noticed. In the framework of our study, it corresponds to the 
superposition of a hard perturbative cut Pomeron interaction with a soft one 
described by a non-cut Pomeron (with complex conjugate contribution, see Fig.1).
This is similar to the superposition of hard and soft interactions which 
characterizes the  SCI approach. We are thus led to propose 
$Disc_2$ as a way to get quantitative predictions, in particular the relative 
normalization of diffractive over non diffractive cross-sections. Indeed,  it 
appears as 
a ``hard'' partonic 
interaction very similar to 
the one describing ordinary deep-inelastic processes, in parallel with a 
``soft''  
correction evolving during a long time, corresponding to the uncut Pomeron 
singularity in the middle graphs (including complex conjugate) of Fig.1. 

The equality between the different discontinuities allows the connection between 
the different models, leading to the results given in section {\bf 1}.

{\centerline{\bf FIGURE}}

\bigskip \input epsf \vsize=8.truecm 
 \epsfxsize=8.cm{%
\centerline{\epsfbox{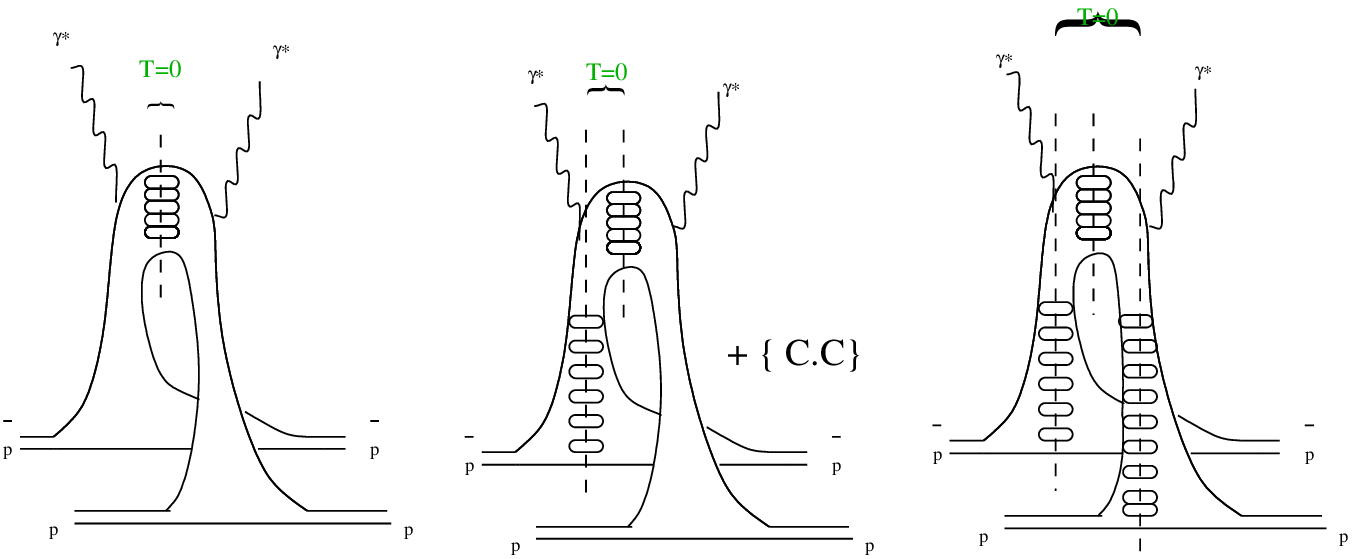}}} {\bf Figure 1}

{\it S-Matrix interpretation of the three approaches to hard diffraction.}
Upper graph: Description of $Disc_1 A(3 \to 3)$; Middle 
graph: Description of $Disc_2 A(3 \to 3)$ and its complex conjugate (candidates 
for the SCI approach); Lower graph: 
Description of $Disc_3 A(3 \to 3)$ (QCD dipole approach).

\bigskip  
\section*{Acknowledgments}
We thank A.Bialas and W.Buchmuller for valuable remarks.

\end{document}